\documentclass[11pt]{article}
\usepackage[hmargin=1in,vmargin=1in]{geometry}
\usepackage[dvipsnames,usenames,table]{xcolor}
\usepackage{hyperref}
\hypersetup{colorlinks=true, urlcolor=Blue, citecolor=Green, linkcolor=BrickRed, breaklinks, unicode}
\usepackage{hchang}

\makeatletter
\renewcommand*{\@fnsymbol}[1]{\ensuremath{\ifcase#1\or *\or \dagger\or \ddagger\or
   \mathsection\or \mathparagraph\or \|\or **\or \dagger\dagger
   \or \ddagger\ddagger \else\@ctrerr\fi}}
\makeatother

\usepackage{url}
\usepackage{graphicx}
\usepackage{caption} 
\usepackage{bbm}
\usepackage{float}
\usepackage{framed}
\usepackage{enumerate}
\usepackage{color}
\usepackage{thm-restate}
\usepackage{comment}

\usepackage{cleveref}
\usepackage{subfiles}

\usepackage{algorithm}
\usepackage[noend]{algpseudocode}
\usepackage{pdfpages}
\usepackage{authblk}

\newtheorem{theorem}{Theorem}

\newtheorem{lemma}{Lemma}

\newcommand{\eps}{\varepsilon}
\advance \topmargin by -\headheight
\advance \topmargin by -\headsep
\evensidemargin \oddsidemargin
\begin{document}

\begin{titlepage}

\title{New weighted additive spanners}
\author[1]{An La\footnote{Email: \href{mailto:anla@umass.edu}{anla@umass.edu}.}}
\author[1]{Hung Le\footnote{Email: \href{mailto:hungle@cs.umass.edu}{hungle@cs.umass.edu}.}}

\affil[1]{University of Massachusetts Amherst}
\date{}

\maketitle

\begin{abstract}
Ahmed, Bodwin, Sahneh, Kobourov, and Spence (WG 2020) introduced additive spanners for \emph{weighted} graphs and constructed (i) a $+2W_{\max}$ spanner with $O(n^{3/2})$ edges and (ii) a $+4W_{\max}$ spanner with $\tilde{O}(n^{7/5})$ edges, and (iii) a $+8W_{\max}$ spanner with $O(n^{4/3})$ edges, for any weighted graph with $n$ vertices. Here $W_{\max} = \max_{e\in E}w(e)$ is the maximum edge weight in the graph.  Their results for $+2W_{\max}$, $+4W_{\max}$, and $+8W_{\max}$ match the state-of-the-art bounds for the unweighted counterparts where $W_{\max} = 1$. They left open the question of constructing a $+6W_{\max}$ spanner with $O(n^{4/3})$ edges. Elkin, Gitlitz, and Neiman (DISC 2021) made significant progress on this problem by showing that there exists a $+(6+\epsilon)W_{\max}$ spanner with $O(n^{4/3}/\epsilon)$ edges for any fixed constant $\epsilon > 0$. Indeed, their result is stronger as the additive stretch is local: the stretch for any pair $u,v$ is $+(6+\epsilon)W_{uv}$ where $W_{uv}$ is the maximum weight edge on the shortest path from $u$ to $v$.

In this work, we resolve the problem posted by Ahmed et al. (WG 2020) up to a poly-logarithmic factor in the number of edges: We construct a $+6W_{\max}$ spanner with $\tilde{O}(n^{4/3})$ edges. 
We extend the construction for $+6$-spanners of Woodruff (ICALP 2010), and our main contribution is an analysis tailoring to the weighted setting. 
The stretch of our spanner could also be made local, in the sense of Elkin, Gitlitz, and Neiman (DISC 2021). We also study the fast constructions of additive spanners with $+6W_{\max}$  and $+4W_{\max}$ stretches. We obtain, among other things, an algorithm for constructing a $+(6+\epsilon)W_{\max}$ spanner of $\tilde{O}(\frac{n^{4/3}}{\epsilon})$ edges in $\tilde{O}(n^2)$ time.    
\end{abstract}
\thispagestyle{empty}

\end{titlepage}

\section{Introduction} \label{sec:intro}

Given $\alpha\geq 1, \beta \geq 0$, an \emph{$(\alpha,\beta)$ spanner} of a (possively edge-weighted) undirected graph $G= (V,E,w)$ is a  spanning subgraph of  $G$, denoted by $H$, such that $d_H(u,v)\leq \alpha \cdot d_G(u,v) + \beta$. If $\beta = 0$, we call $H$ a \emph{multiplicative} spanner with stretch $\alpha$; if $\alpha = 1$, we call  $H$ an \emph{additive} spanner with stretch $+\beta$. 

The notion of (multiplicative) spanner was first formally introduced by Peleg and Schäffer~\cite{PS89}, though its conception appeared earlier~\cite{PU87,PU89}. The multiplicative spanner has been studied extensively; see the recent survey~\cite{ABSHJKS20} and references therein. The basic trade-off between the number of edges and the (multiplicative) stretch is: $(2k-1)$ stretch and $O(n^{1+1/k})$ edges for any $k\geq 1$, see, e.g.,~\cite{ADDJS93}, which is optimal assuming the Erd\H{o}s' girth conjecture~\cite{Erdos64}. Here $n$ denotes the number of vertices of the input graph.

Additive spanners were also studied extensively for \emph{unweighted graphs}. Specifically, the remarkable work of Aingworth, Chekuri, Indyk, and Motwani~\cite{ACIM99} gave a $+2$ spanner with $O(n^{3/2})$ edges. For higher additive stretches, Chechik~\cite{Che13} constructed a $+4$ spanner with   $\tilde{O}(n^{7/5})$\footnote{$\tilde{O}$ notation hides a polylogarithmic factor.} edges; Baswana, Kavitha, Mehlhorn, and Pettie~\cite{BKMP05} constructed a $+6$ spanner with   $O(n^{4/3})$ edges. On the negative side, Abboud and Bodwin showed that $O(n^{4/3})$ edges are almost tight for any constant additive stretch: there exists an $n$-vertex graph such that for any constant $c$ and constant $\epsilon\in (0,1)$, any $+c$ spanner must have $\Omega(n^{4/3-\epsilon})$ edges~\cite{AB17}.  The major open problem in this area is the existence of a $+4$ spanner with $O(n^{4/3})$ edges; solving this problem would complete the stretch-sparsity landscape of additive spanners with constant additive stretch.

Nevertheless,
additive spanners for \emph{weighted graphs} were largely ignored until the recent work of Ahmed et al.~\cite{ABSKS20} who constructed (i) a $+2W_{\max}$ spanner with $O(n^{3/2})$ edges and (ii)  a $+4W_{\max}$ spanner with $\tilde{O}(n^{7/5})$ edges, and (iii)  a $+8W_{\max}$ spanner with $O(n^{4/3})$, for any weighted graph with $n$ vertices. Here $W_{\max} = \max_{e\in E}w(e)$ is the maximum edge weight in the graph. Elkin, Gitlitz, and Neiman~\cite{EGN21} studied a stronger notion of additive stretch, called \emph{$+cW$ spanners}: The stretch between any two vertices $u$ and $v$ is $+cW_{uv}$ where $W_{uv}$ is the maximum edge weight on the shortest path from $u$ to $v$ in $G$. The guarantee $+cW$, referred to as \emph{local error}~\cite{ABHKS21}, is clearly stronger than $+cW_{\max}$. Furthermore, the construction of $+cW$ spanners is often more challenging than $+cW_{\max}$ counterpart. It was observed~\cite{EGN21} that the construction of a $+2W$ spanner with $\tilde{O}(n^{3/2})$ edges was implicit in~\cite{EGN23}.   
As pointed out by Ahmed et al.~\cite{ABSKS20}, results for \emph{unweighted graphs} could not be straightforwardly translated to weighted graphs due to a  technical reason: after adding \emph{arbitrary} $d$ edges per vertex to a spanner, called \emph{$d$-initialization}, a shortest path in \emph{unweighted graphs} missing  $\ell$ edges (in the current spanner) has $\Omega(d\ell)$ neighbors, while the same fact does not hold for \emph{weighted} graphs. They came up with a very simple fix for this in weighted graphs, which requires a clever proof: adding $d$ \emph{lightest} edges per vertex to the spanner, called $d$-light initialization.     The main open problem left in their work is to construct a $+6W_{\max}$ spanner (or more strongly, $+6W$ spanner)  with $O(n^{4/3})$ edges to match with the bound for the $+6$ spanner in unweighted graphs; this is the last missing piece in transferring  major positive results from unweighted graphs  to weighted graphs. Elkin, Gitlitz, and Neiman~\cite{EGN21} come much closer to solving this problem by showing that there exists a $+(6+\eps)W$ spanner with $O(n^{4/3}/\eps)$ edges for any fixed constant $\eps > 0$. However, to get a $+6W$ spanner, the value of $\eps$ must be $1/W_{\max}$, giving  $O(n^{4/3}W_{\max})$ edges which has an undesirable dependency on $W_{\max}$. 
Our first main result is a solution for this problem, as stated in the first item of \Cref{thm:main-6W}.

The technique developed for the first result allows us to extend the construction of $+2$-subsetwise spanners to the weighted setting.
A $+\beta$-subsetwise spanner of a subset $S\subseteq V$ is only required to preserve distances between vertices in $S$ up to additive stretch $+\beta$. 
In unweighted graphs, a $+2$-subsetwise spanner with $O(n\sqrt{|S|})$ edges exists~\cite{BKMP05,Pettie09,CGK13}. 
Ahmed et al.~\cite{ABSKS20} constructed a $+4W_{\max}$-subsetwise spanner with $O(n\sqrt{|S|})$ edges. Elkin, Gitlitz, and Neiman~\cite{EGN21} improved the stretch to $+(2+\epsilon)W$, using $O(n\sqrt{|S|}/\eps)$ edges. 
As shown in the second item of ~\Cref{thm:main-6W}, we obtain a $+2W$-subsetwise spanners with  $\tilde{O}(n\sqrt{|S|})$ edges.


\begin{theorem}\label{thm:main-6W}Let $G = (V,E,w)$ be any undirected, edge-weighted graph with $n$ vertices. 
\begin{itemize}
    \item There exists a $+6W$ spanner for $G$ with $\tilde{O}(n^{4/3})$ edges.
    \item  There exists a $+2W$-subsetwise spanner for any subset $S\subseteq V$ with $\tilde{O}(n\sqrt{|S|})$ edges.
\end{itemize}
\end{theorem}


Our construction for $+6W$ spanners is a one-line adaptation from the algorithm for $+6$ spanners of Woodruff~\cite{Woo10}: we simply change $d$-initialization to $d$-light initialization. Our main technical contribution is in the analysis, which we will provide details in~\Cref{sec:intro_tech_6wspanner}.

The key insight from the construction in~\Cref{thm:main-6W} is that it suffices to preserve the distance between a small subset of pairs, instead of all pairs as in prior works~\cite{ABSKS20, EGN21}. 
This allows us to obtain fast construction of various spanners in weighted graphs.
Our results are summarized in \Cref{thm:time-6W}.

\begin{theorem}\label{thm:time-6W}
Let $G = (V,E,w)$ be any undirected, edge-weighted graph with $n$ vertices.  We can construct for $G$:
\begin{enumerate}
    \item  a $+4W +(2+\epsilon)W_{\max}$ spanner of $\tilde{O}(\frac{n^{4/3}}{\epsilon})$ edges in $\tilde{O}(n^2)$ time. This implies an algorithm for $+(6+\eps)W_{\max}$ spanners of $\tilde{O}(\frac{n^{4/3}}{\epsilon})$ edges in $\tilde{O}(n^2)$ time.
    \item  a $+\max\{6W, 2W_{\max}\}$ spanner with $\tilde{O}(n^{4/3})$ edges in $\tilde{O}(n^{7/3})$ time. This implies an algorithm for $+6W_{\max}$ spanners of  $\tilde{O}(n^{4/3})$ edges  in $\tilde{O}(n^{7/3})$ time. 
    \item a $+6W$ spanner with $\tilde{O}(n^{4/3})$ edges in $\tilde{O}(mn^{2/3})$ time. 
    \item a $+\max\{4W, 2W_{\max}\}$ spanners with  $\tilde{O}(n^{7/5})$ edges in $\tilde{O}( n^{12/5})$ time. This implies an algorithm for $+4W_{\max}$ spanners with  $\tilde{O}(n^{7/5})$ edges in $\tilde{O}(n^{12/5})$ time.
\end{enumerate}
\end{theorem}

In unweighted graphs, no known algorithm can construct a $+\beta$-additive spanners for any constant $\beta$ and run in nearly linear time, despite multiple efforts~\cite{ACIM99,DHZ00, BKMP05, Woo10, Knudsen17, Ald21}. On the other hand, $\tilde{O}(n^2)$ algorithms were known for $+2$~\cite{DHZ00} and $+6$~\cite{BKMP05,Woo10} spanners. For weighted additive spanners, Elkin, Gitlitz, and Neiman~\cite{EGN21} devised an $\tilde{O}(n^2)$-time algorithm for constructing $+2W$ spanners with $\tilde{O}(n^{3/2})$ edges. 
For $+(6+\varepsilon)W$ spanners of $\tilde{O}(n^{4/3})$ edges, their construction runs in $\tilde{O}(mn + n^2)$.
Here we provide several results, achieving 
different time-stretch trade-offs for $+6W$ and $+6W_{\max}$ spanners. 
Most significantly, we obtain a nearly quadratic time algorithm for constructing a  $+(6+\eps)W_{\max}$ spanners with $\tilde{O}(\frac{n^{4/3}}{\epsilon})$ edges. 

For $+4$ spanners in unweighted graphs, careful implementation of Chechik's algorithm~\cite{Che13} gives $O(mn)$ running time, producing a spanner with $\tilde{O}(n^{7/5})$ edges, which is the best-known sparsity bound for $+4$ spanners. Al-Dhalaan~\cite{Ald21} improved the running time to  $\tilde{O}(mn^{3/5})$. The same algorithm could be modified slightly to work for weighted graphs, producing a $+4W + \eps W_{\max}$ spanner in $\tilde{O}(mn^{3/5})$ time.  Here we speed up the construction of Chechik~\cite{Che13} then extend it to the weighted setting, to obtain $+4$ and $+4W_{\max}$ spanners of $\tilde{O}(n^{7/5})$ edges with running time $\tilde{O}(n^{12/5})$. This strictly improves over the algorithm of Al-Dhalaan~\cite{Ald21} when the graph is dense. In particular, when $m = \Theta(n^2)$, our algorithm has running time $\tilde{O}(mn^{2/5})$. Furthermore, the stretch of the spanner by our algorithm does not have the extra term $+\eps W_{\max}$. 

\subsection{Techniques}

\subsubsection{\texorpdfstring{$+6W$ spanners}{\texttwoinferior} }\label{sec:intro_tech_6wspanner}

Baswana, Kavitha, Mehlhorn, and Pettie~\cite{BKMP05} introduced the \emph{path-buying framework} to give the  first construction of a $+6$ spanner with $O(n^{4/3})$ edges in unweighted graphs. However, it is unclear how to extend the path buying framework to weighted graphs while keeping the same bound on the number of edges. 
Knudsen~\cite{Knu14} introduced a simpler greedy algorithm to construct a  $+6$ spanner with the same number of edges: it first adds $n^{1/3}$ edges per vertex to the spanner and then repeatedly adds the shortest path between $u$ and $v$ to the spanner if the additive stretch between them in the current spanners is more than $+6$. The number of edges was then bounded by a very elegant potential function argument. This algorithm has been adapted in different ways to construct spanners for weighted graphs. 

Specifically, Ahmed et al.~\cite{ABSKS20} adapted the algorithm of Knudsen~\cite{Knu14} to construct a $+4W_{\max}$-subsetwise spanner with  $O(n\sqrt{|S|})$ edges for any subset of vertices $S$. Then they used this result along with a standard random sampling technique to derive a  $+8W_{\max}$ spanners with $O(n^{4/3})$ edges. Elkin, Gitlitz and Neiman~\cite{EGN21} on the other hand adapted the algorithm of Knudsen~\cite{Knu14} directly to weighted graphs without going through subsetwise spanners. In particular, after applying a $n^{1/3}$-light initialization, i.e., adding $n^{1/3}$ lightest edges incident to each vertex to the spanner, they sort pairs $(u,v)$ according to $W_{uv}$, and add to the spanner the shortest $u$-to-$v$ path if the current distance in the spanner is more than $+(6+\epsilon)W_{uv}$. The analysis crucially relies on the extra $+\eps W_{uv}$ term to bound the number of edges of the spanner. Specifically, the extra error term translates to an improvement of $\Omega(\epsilon)W_{uv}$ in the distance between $u$ and $v$ every time a shortest path incident to both of them is added. This means once the distance between $u$ and $v$ (in the current spanner) is close enough to $G$, it can only be improved $O(1/\eps)$ times. Applying this argument on top of the potential function argument by  Knudsen~\cite{Knu14} leads to $O(n^{4/3}/\eps)$ bound on the number of edges. It seems that the extra $+\eps W$ is inherent in the technique of Elkin, Gitlitz and Neiman~\cite{EGN21}.

We take a different approach in constructing $+6W$ spanner. Specifically, we directly adapt the random sampling idea of Woodruff~\cite{Woo10}. The algorithm of Woodruff~\cite{Woo10} (for unweighted graphs) first applied a $O(n^{1/3})$-initialization and then handled the stretch by classifying how many edges on shortest paths are missing in the current spanner in an exponential scale. If the number of missing edges is roughly $2^i$ for some $i\in [0,\log n]$, then sampling a set of vertices $D_i$ with probability roughly $O(1/(n^{1/3}2^i))$ and adding to the spanners shortest paths from $D_i$ (to a specific set of randomly sampled vertices) that have at most $2^i$ missing edges. While the main goal of Woodruff~\cite{Woo10} is to design a fast construction of $+6$ spanners in unweighted graphs, we observe that it can be adapted to weighted graphs in a very simple way. All we  need to do is to replace the  $O(n^{1/3})$-initialization by $O(n^{1/3})$-light initialization, a standard procedure used in all weighted additive spanner constructions. One key idea in the stretch analysis is that for every missing edge $(a,b)$ on the shortest $u$-to-$v$ path, we only look at neighbors of $a$ (and of $b$) that are incident to  $O(n^{1/3})$ lightest edges of $a$ (and of $b$) to argue that with high probability, the random sampling will sample a vertex $x$ adjacent to the path and that the edge connecting $x$ and the path has weight at most $W_{uv}$. This way, we can guarantee that the final stretch is $+6W$ (instead of $+6W_{\max}$), as claimed in \Cref{thm:main-6W}.

To construct a $+2W$-subsetwise spanner with $\tilde{O}(n\sqrt{|S|})$ edges, we follow roughly the same technique. In the fast construction of Woodruff~\cite{Woo10} for unweighted graphs, an important subroutine is to find a $+2$-subsetwise spanner with $\tilde{O}(n^{4/3})$ edges for any subset $S$ of size at most $O(n^{2/3})$. We adapt this algorithm for subsets of any size, obtaining the number of edges depending on the size of $S$. Again, we employ  $O(n^{1/3})$-light initialization and then randomly sample the sets $\{D_i\}_{i=0}^{\log(n)}$ as above. Then for each $D_i$, we add shortest paths with roughly $2^{i}$ missing edges from $D_i$ to $S$. We use the same argument for the $+6W$ spanner to show that the stretch is $+2W$, completing \Cref{thm:main-6W}. 

\subsubsection{Fast construction}\label{sec:intro_fast}

To obtain fast construction, we need a subroutine that efficiently finds shortest paths from  sets of randomly sampled vertices. In designing a nearly quadratic time algorithm for unweighted graphs, Woodruff~\cite{Woo10} introduced a subroutine called \emph{fewest heavy edges with almost shortest walks} (FHEASW). This subroutine finds, for any given two vertices $u$ and $v$, a minimum number of \emph{heavy edges},  those that are incident to very high degree vertices, on paths of lengths in $\{d_G(u,v), d_G(u,v) + 1, d\ldots, d_G(u,v)+4\}$. The subroutine is implemented via a dynamic program that tracks both vertices and the additive stretch $i \in \{0,\ldots, 4\}$. Generalizing this subroutine to weighted graphs, one has to keep track of every additive stretch $i \in \{0,1,\ldots, 4W_{\max}\}$, leading to $O(mW_{\max})$  running time. To improve this running time, we borrow an idea by Al-Dhalaan~\cite{Ald21}: reweighting the edges of the graph and then applying Dijkstra.  (Al-Dhalaan~\cite{Ald21} used this idea for constructing $+4$ and $+(4+\eps)W$ spanners, and while we construct $+6W$ spanners, which leads to several other differences in the implementation.) Specifically, we  increase the weights of missing edges of $G$ by a small amount of roughly $\frac{\eps}{g}W_{\max}$, where $g$ is the constraint on the number of missing edges. The weight increase guarantees that the shortest path found (after restoring the original weights) still has a good additive stretch while the number of missing edges is only a constant time worst than desired.  
Due to the reweighting step, the final additive stretch has an extra $+\eps W_{\max}$. But the path-finding now has running time $\tilde{O}(m)$ instead of $O(mW_{\max})$, which ultimately leads to an overall running time of $\tilde{O}(n^2)$, proving Item 1 in \Cref{thm:time-6W}. 

For the algorithm in Item 2 of \Cref{thm:time-6W}, we take a different (and simpler) approach. Here we could not tolerate an amount of $+\eps W_{\max}$ error in the stretch and hence reweighting is not applicable. We instead develop a subroutine called  \emph{missing edges constrained shortest path} (MECSP) that combines dynamic programming and Dijkstra algorithms. Specifically, we keep track of the number of missing edges in the shortest path from the root vertex to a vertex $v$ in a table entry $f[\ell,v]$, where $\ell$ is the number of missing edges, and then apply Dijkstra's ideas to update $f[\ell,v]$ from the neighbors of $v$. The running time is $\tilde{O}(\ell m)$. However, we only apply this subroutine from a small subset of vertices of size roughly $O(n^{2/3}/\ell)$.
In addition, we have a pre-processing step so that the subroutine runs in a sub-graph of at most $O(n^{5/3})$ edges.
This step costs $\tilde{O}(mn^{1/3})$, thus the total running time is 
$\tilde{O}(mn^{1/3}) + \tilde{O}(\ell n^{5/3})\cdot O(n^{2/3}/\ell) = \tilde{O}(n^{7/3})$ as claimed in Item 2 of \Cref{thm:time-6W}.

We apply the same idea (of using MECSP and pre-processing step) to speed up the construction of $+4$ spanners of Chechik~\cite{Che13}, improve the running time to $\tilde{O}(n^{12/5})$. 
Replacing $d$ initialization by $d$-light initialization, we immediately obtain an algorithm with running time $\tilde{O}(n^{12/5})$ for $+4W_{\max}$ spanners of $\tilde{O}(n^{7/5})$ edges, as stated in Item 4 of~\Cref{thm:time-6W}.

Given the MECSP, we could directly plug it into our randomized algorithm showing the existence of a $+6W$ spanner with $\tilde{O}(n^{4/3})$ edges. In particular, we run  MECSP from each vertex $v$ in the sampling set $D_i$, obtaining the total running time of $\tilde{O}(mn^{2/3})$ as claimed in Item 3 of \Cref{thm:time-6W}.

\section{Preliminaries}
    Let $\pi$ be a path in $G$. Let $\pi(u, v)$ be a sub-path from $u$ to $v$ of $\pi$.
    We denote by $|\pi|$ the total weights of edges in $\pi$, and by $W_{uv}$ the maximum weight of edges in the shortest path from $u$ to $v$. 
    Let \emph{$N(v)$} be neighbors of $v$ in $G$, and \emph{$N[v] = N(v) \cup \{v\}$}.  For a set $S \subseteq V$ and a path $\pi$, it is defined as \emph{$N[S] = \cup_{v \in S} N[v]$}, and  
    $N[\pi] = \cup_{v \in \pi} N[v]$.
    
    $H$ is \emph{$d$-light initialization} of $G$ if it is obtained by adding $d$ lightest-weight edges of each vertex. 
    A vertex is \emph{$d$-heavy} if its degree is at least $d$.
    \emph{Missing edges} are edges that are in $G$ but not in $H$.

    In this paper, we use the following lemmas:

    \begin{restatable}{lemma}{randomsample}
    \label{lmm:rdm_smpl}
    Given an undirected graph $G$ of $n$ vertices:
    \begin{enumerate}
        \item Let $S$ be a set of  randomly sampled vertices with probability $\frac{2\ln{n}}{d}$. For every $d$-heavy vertex $u$, $N[u] \cap S \neq \emptyset$ with a probability at least $1 - \frac{1}{n}$.
        \item Let $S$ be a set of randomly sampled vertices with probability $\frac{\ln{n}}{k}$. 
        For every path $\pi$, if $|N[\pi]| = \Omega(k)$, then $N[\pi] \cap S \neq \emptyset$ with a probability at least $1 - \frac{1}{n}$.
    \end{enumerate}
    \end{restatable}

    \begin{lemma}[~\cite{ABSKS20, EGN21}]\label{lmm:hittingset}
    Suppose that $H$ is a $d$-light initialization of an undirected weighted graph $G$. 
    Let $\pi$ be the shortest path between $s$ and $t$ with $l$ missing edges. 
    Then there is a set $S \subseteq V$ such that:
    \begin{enumerate}
        \item $|S| = \Omega(dl)$.
        \item For any $v \in S$, there is a vertex $r \in \pi$: $(r, v) \in E(H)$ and $w(r, v) \leq W_{st}$.
    \end{enumerate}
    \end{lemma}

    The following lemma is implicit in the work of Elkin, Gitlitz, and Neiman~\cite{EGN21}:

    \begin{restatable}{lemma}{vertexhit}
    \label{lmm:vertexhit}
            Suppose that $H$ is a $d$-light initialization of an undirected weighted graph $G$. Given a shortest path $\pi$ between $s$ and $t$,
            let $D$ be a set of randomly sampled vertices with probability $\frac{2\ln{n}}{d}$. For any $d$-heavy vertex $u \in \pi$, there exists $(u, v) \in E(H)$ such that $v \in D$ and $w(u, v) \leq W_{st}$ with a probability at least $1 - 1/n$.
    \end{restatable}

    The proof of~\Cref{lmm:rdm_smpl} and~\Cref{lmm:vertexhit} are simple and reproduced  in Appendix.
    We have the following claim, which is crucial to construct $+6W_{\max}$ or $+6W$ spanners.
    \begin{lemma}\label{lmm:pathhit}
        Suppose that $H$ is a $d$-light initialization of an undirected weighted graph $G$. 
        Let $\pi$ be the shortest path between $s$ and $t$ in $G$ with $l$ missing edges.
        If $R$ is a set of randomly sampled vertices with probability $\frac{\ln{n}}{dl}$, then there is an edge $(r, v) \in E(H)$ such that $r \in \pi, v \in R$ and $w(r, v) \leq W_{st}$ with a probability at least $1 - 1/n^c$ where $c$ is a constant.
    \end{lemma}
    \begin{proof}
        Create $R$ by picking each vertex with probability $\frac{\ln{n}}{dl}$. 
    By~\Cref{lmm:hittingset}, there is a set $S$ with size $\Omega(dl)$, and every vertex $v\in S$ has an edge $(v, r) \in E(H)$ such that $w(r, v) \leq W_{st}$. 
    Suppose that $|S| = cdl$ where $c$ is a constant.
    Then:
\begin{equation*}
        P[S \cap R = \emptyset] = (1 - p_R)^{|S|} 
        \leq \left(1 - \frac{\ln{n}}{dl}\right)^{cdl} = \frac{1}{n^c}
\end{equation*}
    Thus $P[S \cap R \neq \emptyset] \geq 1 - \frac{1}{n^c}$.
    \end{proof}

	\section{\texorpdfstring{$+6W$ spanners}{\texttwoinferior} }\label{sec:6ws}

    \begin{figure}[t!]
        \begin{minipage}{.60\textwidth} \centering      \includegraphics{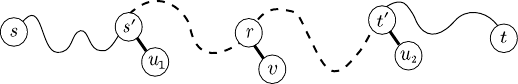}
         \end{minipage}
        \hfill\begin{minipage}{.40\textwidth} \centering
        \includegraphics {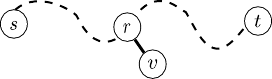}
        \end{minipage}
        \caption{
        Illustration of a path between $s$ and $t$ after $d$-light initialization,
        the left figure is for~\Cref{alg:6WS} in~\Cref{sec:6ws} ($+6W$ spanners), the right figure is for the algorithm in~\Cref{sec:2wss} ($+2W$-subsetwise spanners). 
        Solid lines are sub-paths with no missing edge, the dash lines are the sub-paths with some missing edges. 
        Each thick and short line is an edge added to $H$.
        }
        \label{fig:enter-label}
    \end{figure}

    In this section, we prove Item 1 of \Cref{thm:main-6W}.
    The construction for $+6W$ spanners is as follows: Let $H$ be a $d$-light initialization of $G$ where $d = n^{1/3}$.
    First, create $R$ by randomly sampling vertices with probability $p_R =\frac{2\ln{n}}{d}$. 
	Let $i$ be an integer in the range $[0 \ldots \log(n)]$. 
	For each value of $i$, create a set $D_i$ by randomly sampling vertices with probability $p_i$, where $p_i = \frac{\ln{n}}{d2^i}$.
    Then for each pair $u, v$, among paths from $u$ to $v$ that have less than $2^{i+1}$ missing edges, add to $H$ the shortest one.

    \begin{algorithm}
    \caption{$+6W$ Spanners}\label{alg:6WS}
    \begin{algorithmic}[1]
    \Procedure{Spanner6W}{$G = (V, E, w)$}
        \State$d\gets n^{1/3}$
        \State$H\gets d$-light initialization of $G$
        \State$R\gets$ randomly sampling vertices with probability $p_R = \frac{2\ln{n}}{d}$
        \For{$i\gets [1..\log(n)]$}
            \State$D_i\gets$  randomly sampling vertices with probability $p_i= \frac{\ln{n}}{d2^i}$
            \For {$v \in D_i$}
                \For {$u \in R$}
                    \State $\mathcal{P}_{uv}\gets \{ \textrm{path $\pi$ from $v$ to $u$} \mid \pi \textrm{ has less than $2^{i+1}$ missing edges}\}$
                    \If {$\mathcal{P}_{uv}\neq \emptyset$}
                    \State $\pi^* \gets \arg\min_{\pi\in \mathcal{P}_{uv}} |\pi|$
                    \State $H\gets H\cup \pi^*$
                    \EndIf
                \EndFor
            \EndFor
        \EndFor
        \State \textbf{return} $H$
    \EndProcedure
    \end{algorithmic}
    \end{algorithm}
	
	\begin{theorem}\label{thm:6W_stretch_size}
	\Cref{alg:6WS} returns a $+6W$ spanner of $\tilde{O}(n^{4/3})$ edges. 
	\end{theorem}

	\begin{proof}
	    Firstly, we prove that the algorithm returns $+6W$ spanners.
	    Consider the shortest path $\pi$ of a pair $(s, t) \in V \times V$.
        If $\pi$ does not have a missing edge after the initialization, then $\pi$ is preserved in $H$.
        Otherwise, $\pi$ has at least two $d$-heavy vertices. 
	    Let $s', t'$ be endpoints of missing edges that are respectively closest to $s$ and $t$. 
        It follows that $s'$ and $t'$ are $d$-heavy vertices, or each of $s'$ and $t'$ has at least $d$ neighbors in $H$. 
	    By \Cref{lmm:vertexhit}, there exist $u_1, u_2 \in R$ such that the edges $(s', u_1), (t', u_2)$ are in $H$ and $w(u_1, s') \leq W_{st}$, $w(u_2, t') \leq W_{st}$.
	    
	    Let $l$ be the number of missing edges of $\pi$. Suppose $l \in [2^i, 2^{i+1})$ for an integer $i\in [0\ldots \log(n)]$. 
	    By \Cref{lmm:pathhit}, there exist $v \in D_i$ and $r \in \pi$ such that $(r, v) \in E(H)$ and $w(r, v) \leq W_{st}$.
	    See Figure 1 for an illustration.
	    
	    By lines 7 to 12 of the algorithm, since $u_1 \in R$ and $v \in D_i$, there is a path between these vertices in $H$.
        Observe that $\pi(s', r)$ has at most $l < 2^{i+1}$ missing edges.
        Consider the path $\pi' = (u_1, s') \odot \pi(s', r) \odot (r, v)$, which walks on the edge $(u_1, s')$, the sub-path $\pi(s', r)$, and the edge $(r, v)$.
        $\pi'$ also has at most $l$ missing edges, thus $\pi'$ is considered by the algorithm when finding a path between $u_1$ and $v$, i.e. $\pi' \in \mathcal{P}_{u_1v}$. It implies:
	    \begin{equation*}\begin{aligned}
	        d_H(u_1, v) &\leq |\pi'| = w(u_1, s') + d_G(s', r) + w(r, v) 
	        \\
	       &\leq d_G(s', r) + 2W_{st} 
	    \end{aligned}\end{equation*}

	    Similarly, the path $\pi'' = (u_2, t') \odot \pi(t', r) \odot (r, v)$, which walks on the edge $(u_2, t')$, the sub-path $\pi(t', r)$, and the edge $(r, v)$, is considered by the algorithm when finding a path from $u_2$ to $v$. Therefore: 	    
        \begin{equation*}
        \begin{aligned}
            d_H(u_2, v) &\leq |\pi''| = w(u_2, t') + d_G(t', r) + w(r, v) 
	        \\
            &\leq d_G(t', r) + 2W_{st} 
        \end{aligned}    
        \end{equation*}
        By triangle inequality, $d_H(u_1, u_2) \leq d_H(u_1, v) + d_H(u_2, v)$, which is at most $d_G(s', r) + 2W_{st} + d_G(t', r) + 2W_{st}$. 
        Since $r \in \pi(s', t')$, we obtain $d_H(u_1, u_2) \leq d_G(s', t') + 4W_{st}$.
        By definition of $s'$ and $t'$, all edges in $\pi(s, s')$ and $\pi(t', t)$ are added to $H$ by $d$-light initialization, or $d_H(s, s') = d_G(s, s')$ and $d_H(t', t) = d_G(t', t)$.
        We obtain:
        \begin{equation*}\begin{aligned}
	        d_H(s, t) &\leq  d_H(s, s') + w(u_1, s') + d_H(u_1, u_2) + w(u_2, t') + d_H(t', t) \\
	       &\leq d_G(s, s') + W_{st} + (d_G(s', t') + 4W_{st}) +  W_{st} + d_G(t', t) \\
	       &= d_G(s, t) + 6W_{st}
	    \end{aligned}\end{equation*}

    Now we bound the size of spanners returned by the algorithm. 
    $d$-light initialization requires $O(nd)$ edges. After that, for each value of the integer $i$, the algorithm adds the shortest path between $R$ and $D_i$ and these paths have less than $2^{i+1}$ missing edges. 
    The expected size of $D_i$ is $\tilde{O}(\frac{n}{d2^i})$, thus the expected number of missing edges added is $O\left(|R||D_i|2^{i+1}\right) = \tilde{O}\left(\dfrac{n^2}{d^2}\right)$. 
    The total edges for all values of $i \in [0 \ldots \log(n)]$ is therefore $\tilde{O}\left(\dfrac{n^2}{d^2}\right)$. 
    With $d = n^{1/3}$, we conclude that $H$ has $\tilde{O}(n^{4/3})$ edges. 
	\end{proof}

\section{\texorpdfstring{$+2W$-subsetwise spanners}{\texttwoinferior}}\label{sec:2wss}
    In this section, we show the details of Item 2 of \Cref{thm:main-6W}.
    The construction of $+2W$-subsetwise spanners is similar to $+6W$ spanners. First let $H$ be the $d$-light initialization of $G$, where $d = \sqrt{|S|}$. Let $i$ be an integer in the range $[0\ldots \log(n)]$. For each value of $i$, create $D_i$ by randomly sampling vertices with probability $p_i = \frac{\ln{n}}{d2^{i}}$. Then for each pair $(v, u)$ where $v \in D_i$ and $u \in S$, consider all paths that have less than $2^{i+1}$ missing edges, add to $H$ the shortest among them.

	\begin{theorem}
        The algorithm returns $+2W$-subsetwise spanner of $\tilde{O}(n\sqrt{|S|})$ edges.	
	\end{theorem}
	
	\begin{proof}
	    Consider the shortest path $\pi$ of a pair $(s, t) \in S \times S$.
        If $\pi$ does not have a missing edge after $d$-light initialization, it is preserved in $H$.
        Otherwise, 
	    suppose that $l$ is the number of missing edges of $\pi$ and $l \in [2^i, 2^{i+1})$ for an integer $i \in [0\ldots \log(n)]$. 
	    By \Cref{lmm:pathhit}, there exists $r \in \pi$ and $v \in D_i$ such that $(r, v)$ is in $H$ and $w(r, v) \leq W_{st}$. See Figure 1 for an illustration.
	      
        Since $v \in D_i$ and $s \in S$, the algorithm adds to $H$ a path between $s$ and $r$ with less than
        $2^{i+1}$ missing edges.    
        Let $\pi'$ be the path from $s$ to $v$ which walks along $\pi(s, r)$ then traverses the edge $(r, v)$, i.e. $\pi' = \pi(s, r) \odot (r, v)$. 
	    Observe that $\pi'$ has less than $2^{i+1}$ missing edges, thus $d_H(s, v) \leq |\pi'| = d_G(s, r) +  w(r, v)$.
	    Similarly, we have $d_H(t, v) \leq d_G(t, r) + w(r, v)$.
	    Thus: 
	    \begin{equation*}\begin{aligned}
	        d_H(s, t) &\leq d_H(s, v) + d_H(v, t) \quad & \textrm{(by triangle inequality)} \\
	        &\leq d_G(s, r) + w(r, v) + d_G(t, r) + w(r, v) \quad & \textrm{(by the algorithm)}\\
	        &\leq d_G(s, t) + 2w(r, v) \quad & \textrm{(since $r \in \pi$)}\\
	        &\leq d_G(s, t) + 2W_{st} \quad & \textrm{}
	\end{aligned}\end{equation*}

        Now we bound the size of $H$.
	    The $d$-light initialization of $G$ requires $O(nd) = O(n\sqrt{|S|})$ edges. For each value $i \in [0\ldots \log(n)]$ the algorithm adds shortest paths  between $S$ and $D_i$, which adds $O(|S||D_i|2^{i+1})$ missing edges. The expected size of $D_i$ is $\tilde{O}(\frac{n}{d2^i})$ vertices, thus the expected number of missing edges added is $\tilde{O}(|S|\frac{n}{d2^i}2^{i+1}) = \tilde{O}(n\sqrt{|S|})$ edges. 
	    We conclude that the expected size of $H$ is $\tilde{O}(n\sqrt{|S|})$.
	\end{proof}

    \section{Fast construction for weighted additive spanners}
    \label{sec:fast_cons}
     Algorithms in~\Cref{sec:6ws} and~\Cref{sec:2wss} have a step of finding $\mathcal{P}_{uv}$, a set of all paths between $u$ and $v$ for $u \in R$ and $v \in D_i$ with a constraint on the number of missing edges. Then we add to $H$ the shortest path $\pi^*$ in $\mathcal{P}_{uv}$: $\pi^* \gets \arg\min_{\pi_0\in \mathcal{P}_{uv}} |\pi_0|$.
    Here we provide efficient ways to handle these steps to obtain fast construction for variants of $+6W$ spanners, and the same technique is applied to $+2W$-subsetwise and $+4W_{\max}$ spanners. 

    Observe that we add to $H$ shortest paths from vertices of random samples $D_i$.
    Furthermore, by the analysis of stretch above, after $d$-light initialization, there is a ``candidate'' path $\pi'$ such that: (i) $\pi'$ contains vertices in $D_i$; (ii) $\pi'$ has the same number of missing edges as $\pi$, the shortest path between $s$ and $t$ we want to preserve approximately; (iii) $|\pi'|$ is not too large compared to $|\pi|$. 
    Therefore, we need a subroutine to find $\pi^*$ such that it is as ``good'' as $\pi'$: $\pi^i$ has no more than $2^{i+1}$ missing edges and $|\pi^*| \leq |\pi'|$.
    More importantly, the analysis shows that using $\pi^*$ is sufficient to get the desired stretch.

    There are two ways to find $\pi^*$ from $\pi'$.
    The first one is minimizing the weights of paths while keeping track of the number of missing edges at the same time.
    The second way is re-weighting edges before applying Dijkstra.    
    Each of them is developed as a subroutine, MECSP and Weak CSSSP respectively.

    Previously,   
    Woodruff had a similar idea for unweighted graphs and developed a subroutine called fewest heavy edges with almost shortest walks (FHEASW)~\cite{Woo10} to find a path with the minimum number of missing edges and the additive stretch is no more than $+c$ comparing to $\pi'$. 
    Since $c$ is a constant, FHEASW has the running time of $O(cm) = O(m)$ where $m$ is the number of edges of the input graph. 
    Now with weighted graphs, the additive term is $+cW_{\max}$ instead of $+c$, thus the running time becomes $O(mW_{\max})$.
    To remove the dependency on $W_{\max}$, we design a subroutine called missing edge constrained shortest paths (MECSP).
    We refer back to~\Cref{sec:intro_fast} for the main ideas of FHEASW and MECSP.
    The details of MECSP are shown in~\Cref{sec:subroutine}.

    For the second way to find $\pi^*$ from $\pi'$, re-weighting edge is simply 
    adding a small value to each missing edge. 
    If this amount is too small, shortest paths returned by Dijkstra could have many missing edges. 
    Meanwhile, a too-high value induces ``noise'': we are finding paths with the smallest number of missing edges rather than the original weight.
    The extra amount of a missing edge is defined appropriately by considering the candidate path. 
    Al-Dhalaan~\cite{Ald21} applied this idea to develop a subroutine called weak constrained single-source shortest paths (Weak CSSSP).
    The goal of the author was to speed up the construction of $+4$ and $+4W$ spanners.
    We simplify the technique of re-weighting edge to have fast construction for $+4$, $+4W$ and $+6W$ spanners.
    The simplification is presented in~\Cref{sec:subroutine}.
    
    Applying different subroutines, we obtain different results.
    In addition to finding $\pi^*$ efficiently, we also apply another speeding-up technique~\cite{ACIM99, DHZ00, Woo10}, which handles vertices differently depending on their degrees. 
    Specifically, we have a pre-processing step for high-degree vertices, then run our path-finding subroutines on the graph of low-degree vertices only.
    By using re-weighting edges and MECSP mentioned above, 
    we respectively obtain Item 1 and Item 2 of~\Cref{thm:time-6W}, which are shown in~\Cref{sec:detail_fast1} and~\Cref{sec:detail_fast2}.
    To have the result as Item 3 of the theorem, in~\Cref{sec:detail_fast3} we simply apply MECSP to find paths added to $H$, without the pre-processing step.  
    Finally, in~\Cref{sec:detail_fast4} we apply these techniques to speed up the construction of $+4$ and $+4W_{\max}$ spanners to derive the Item 4 of~\Cref{thm:time-6W}. 
    \subsection{Subroutines}\label{sec:subroutine}
    \paragraph*{Weak Constrained Single-Source Shortest Paths.~}
    In weak constrained single-source shortest paths (Weak CSSSP)~\cite{Ald21}, we reweight the edges of the graph and then apply Dijkstra.
    For each missing edge, we add a small amount of weight: $\frac{\epsilon_0}{g}W_{\max}$, given $g$ is the maximum number of missing edges and $\epsilon_0 \in (0, 1)$. Let $\phi(\pi)$ be the number of missing edges of $\pi$. 
    The total weight of $\pi$ after reweighting is:
    \begin{equation*}\begin{aligned}
        \delta(\pi) = |\pi| + \phi(\pi) \dfrac{\epsilon_0}{g}W_{\max}
    \end{aligned}\end{equation*}

    The analysis of Weak WCSSP in the work of Al-Dhalaan~\cite{Ald21} was specific for $+4$ and $+4W$ spanners (for $+4$ spanners, all edges have unit weights and $W_{\max} = 1$). We adjust it so that it can be applied for any additive stretch as follows: 
    
    \begin{lemma}\label{lmm:wcssp}
        Given $\varepsilon_0 \in (0, 1)$ and a shortest path $\pi$ going from $s$ to $t$ where $\pi$ has less than $l$ missing edges.
        If there exists a path $\pi'$ such that:
        \begin{itemize}
            \item $|\pi'| \leq |\pi| + c_1W_{st} + c_2W_{\max}$, where $c_1, c_2$ are small integers.
            \item  $\pi'$ has less than $l$ missing edges: $\phi(\pi') < l$.  
        \end{itemize} 
        then Weak CSSSP returns a path  $\pi^*$ going from $s$ to $t$ such that:
    \begin{itemize} 
        \item $\pi^*$ is near-optimal: $|\pi^*| < |\pi| + c_1W_{st} + (c_2 + \epsilon_0)W_{\max}$. 
        \item $\pi^*$ satisfies the weak constraint: $\pi^*$ has less than $\dfrac{kl}{\epsilon_0}$ missing edges.
    \end{itemize}
    \end{lemma}

    \begin{proof}
        Setting $g = l$, let $\pi^*$ be the path going from $s$ to $t$ returned by Weak CSSSP.
    
        To prove $\pi^*$ is near-optimal, by definition of $\delta$ we know $|\pi^*| \leq \delta(\pi^*)$, and by $\phi(\pi') < l$ we have $\delta(\pi') < |\pi'| + \epsilon_0W_{\max}$. By Dijkstra, $\delta(\pi^*) \leq \delta(\pi')$.
        It follows that:
        \begin{equation*}\begin{aligned}
            |\pi^*|  &\leq \delta(\pi^*)\leq \delta(\pi') < |\pi'| + \epsilon_0 W_{\max} \\
            &\leq |\pi| + c_1W_{st} + (c_2+\epsilon_0)W_{\max}
        \end{aligned}\end{equation*} 

        To prove the weak constraint, for contradiction, suppose $\phi(\pi^*) > \dfrac{kl}{\epsilon_0}$. Then $\delta(\pi^*) > |\pi^*| + kW_{\max}$. $\pi$ is the shortest path, thus $|\pi^*|\geq |\pi|$ and $\delta(\pi^*) > |\pi| + kW_{\max}$. On the other hand, we know that $\delta(\pi') < |\pi'| + \epsilon_0W_{\max}$. Since $|\pi'| \leq |\pi| + c_1 W_{st} + c_2W_{\max}$,  we obtain
        $\delta(\pi') < |\pi| + c_1W_{st}+ (c_2+\epsilon_0)W_{\max}
        $. Setting $k = c_1 + c_2 + 1$, we have $\delta(\pi^*) > \delta(\pi')$, which contradicts to Dijkstra. 
    \end{proof}

    \paragraph*{Missing Edge Constrained Shortest Paths.~}

    Let MECSP be a subroutine that is given a graph $G$, a source vertex $u$ and an integer $g$; then returns the shortest paths starting from $u$ to every $v \in V$ with the number of missing edges less than $g$.
    The subroutine runs Dijkstra $g$ times and maintains a dynamic tables $f$: $f[l, v]$ is the minimum weight of paths from $u$ to $v$ with at most $l$ missing edges. While updating the table, keep track of vertex which minimizes $f[l, v]$ to obtain the corresponding  shortest path from $u$ to $v$.

    Consider a fix value of $l$. Let $S$ be a set of vertices such that their distances to $u$ are minimized. Similar to Dijkstra, we find a new vertex $v \not\in S$ that has the minimum distance from $u$, then add $v$ to $S$. After that, we update the distances (from $u$) to every $t \in N(v) \setminus S$ by considering the edge $(t, v)$. If $(t, v)$ is a missing edge, $f[l, t]$ is compared with $f[l-1, v] + w(t, v)$. Otherwise, it is compared with $f[l, v] + w(t, v)$. The details of MECSP is as follows:
    
    \begin{algorithm}
    \caption{ Missing Edge Constrained Shortest  Paths}\label{alg:MECSP}
    \begin{algorithmic}[1]
    \Procedure{MECSP}{$G = (V, E, w), u, g$}
    \State Given graph $G$, vertex $u$, constraint number of missing edges $g$
    \State Apply Dijkstra to compute $f[0, v] \quad \forall v \in V$ 
    \For {$l \gets [1, g-1]$}
        \State $S\gets\emptyset$, $f[l, u] \gets 0$
        \For {$v\in V$}    
        \State $f[l, v] \gets \infty$
        \EndFor
        \While {$|S| < |V|$}
            \State $v \gets \argmin_{t \not\in S} f[l, t]$
            \State $S \gets S \cup \{v\}$
            \For {$t \in N(v) \setminus S$} 
            \State $f[l, t] \gets \min\{f[l-1, t], f[l, t]\}$
            \If {$(t, v) \in H$}
                \State $f[l, t] = \min\{f[l, t], f[l, v] + w(t, v)\}$
            \Else
                \State $f[l, t] = \min\{f[l, t], f[l-1, v] + w(t, v)\}$
            \EndIf
            \EndFor
        \EndWhile
    \EndFor
    \State Track $f$ to return paths from $u$ to every $v \in V$.
    \EndProcedure
    \end{algorithmic}
    \end{algorithm}

   \subsection{ \texorpdfstring{$+6W$ spanners in $\tilde{O}(mn^{2/3})$ time}{\texttwoinferior} }
    \label{sec:detail_fast3}

    Here is the details of construction $+6W$ spanners of $\tilde{O}(n^{4/3})$ edges in $\tilde{O}(mn^{2/3})$ time, as claimed in Item 3 of \Cref{thm:time-6W}.
    This is the~\Cref{alg:6WS} with using MECSP.  

   \begin{algorithm}
    \caption{Fast construction for $+6W$ spanners}\label{alg:6W_Fast}
    \begin{algorithmic}[1]
    \Procedure{FastSpanner6W}{$G = (V, E, w)$}
        \State$d\gets n^{1/3}$
        \State$H\gets d$-light initialization of $G$
        \State$R\gets$ randomly sampling vertices with probability $p_r = \frac{2\ln{n}}{d}$
        \For{$i\gets [1..\log(n)]$}
            \State$D_i\gets$  randomly sampling vertices with probability $p_i= \frac{\ln{n}}{d2^i}$
            \For {$v \in D_i$}
                \State $\mathcal{P}_{v} \gets \Call{MECSP}{G, v, 2^{i+1}}$
                // contains paths returned by MECSP from $v$ to all vertices
                
                \For {$u \in R$}
                    \State $\pi^* \gets$ the path from $v$ to $u$ in $\mathcal{P}_v$ 
                    \State $H\gets H\cup \pi^*$
                \EndFor
            \EndFor
        \EndFor
        \State \textbf{return} $H$
    \EndProcedure
    \end{algorithmic}
    \end{algorithm}
    
 \begin{theorem}
	\Cref{alg:6W_Fast} returns a $+6W$ spanner of $\tilde{O}(n^{4/3})$ edges of $G$ in $\tilde{O}(mn^{2/3})$ time.
	\end{theorem}

    \begin{proof}
        The analysis of stretch and size is the same as in \Cref{thm:6W_stretch_size}. 
        
        Now we compute the running time. 
        $d$-light initialization takes $O(m)$ time.
        We call MECSP for each vertex in $D_i$. The running time of MECSP is $\tilde{O}(m2^i)$ and the expected size of $D_i$ is $\tilde{O}(\frac{n}{d2^i})$. Thus, the running time of MECSP for all vertices in $D_i$ is $\tilde{O}(|D_i|m2^i) = \tilde{O}(\frac{n}{d2^i}m2^i) = \tilde{O}(\frac{nm}{d})$. With $d = n^{1/3}$, it is $\tilde{O}(mn^{2/3})$. We have $\log{n}$ different sets $D_i$, thus the total running time is $\tilde{O}(mn^{2/3})$ in expectation.  
    \end{proof}
 
    \subsection{ \texorpdfstring{$+\max\{6W, 2W_{\max}\}$ spanners in $\tilde{O}( n^{7/3})$ time}{\texttwoinferior} }
    \label{sec:detail_fast2}

    In this section, we modify \Cref{alg:6W_Fast} to obtain a faster construction, as the result of Item 2 \Cref{thm:time-6W}.
    This step finds $+2W_{\max}$ spanners for shortest paths that contain $\mathcal{E}$-heavy vertices. After that, we remove all edges of $\mathcal{E}$-heavy vertices and apply the original algorithm.
    
    \begin{algorithm}
    \caption{Fast construction for $+6W_{\max}$ spanners}\label{alg:6WMax_Fast}
    \begin{enumerate}
        \item Filtering edges:
        
        $\mathcal{E} \gets n^{2/3}$
        
        $S$ $\gets$ randomly sampling vertices with probability $\frac{2\ln{n}}{\mathcal{E}}$
        
        Add to $H$ the shortest-path tree from each vertex of $S$ to all vertices of $V$ 
        
        Remove all edges of $\mathcal{E}$-heavy vertices out of $G$.
        
        \item \Call{FastSpanner6W}{G}
    \end{enumerate}        
    \end{algorithm}
    
 \begin{theorem}
	\Cref{alg:6WMax_Fast} returns a $+\max\{6W, 2W_{\max}\}$ spanner of $\tilde{O}(n^{4/3})$ edges of $G$ in $\tilde{O}(n^{7/3})$ time.
	\end{theorem}
 
	\begin{proof}

    By~\Cref{thm:6W_stretch_size}, we know the expected number of edges added by step 2 is $\tilde{O}(n^{4/3})$, and $d_H(s, t) \leq d_G(s, t) + 6W_{st}$ for some pairs $(s, t)$ preserved by this step.
    It is sufficient to consider the filter-edge step only.
    $E[|S|] = \tilde{O}(\frac{n}{\mathcal{E}})$, thus the expected number of edges added is $O(|S|n) =\tilde{O}(\frac{n^2}{\mathcal{E}})$, which is $\tilde{O}(n^{4/3})$.
    With the stretch, consider a path $\pi$ between $s$ and $t$. If there is no $\mathcal{E}$-heavy vertex in $\pi$, it is preserved by step 2. Otherwise, let $r \in \pi$ be $\mathcal{E}$-heavy vertex. By the first item of \Cref{lmm:rdm_smpl}, there exists $v \in S$ such that $(v, r)$ is in $G$. Thus, there is a path walking from $s$ to $u$ and $u$ to $t$ in $H$:  
    \begin{equation*}\begin{aligned}
        d_H(s, t) &\leq d_H(s, v) + d_H(v, t) 
        &\textrm{(triangle inequality)} \\
        &= d_G(s, v) + d_G(v, t) 
        &\textrm{(Add shortest-path tree from $v \in S$ to $V$)} \\
        &\leq  d_G(s, v) + w(r, v) + d_G(r, t) + w(r, v) 
        &\textrm{(triangle inequality)} \\
        &= d_G(s, t) + 2w(r, v) 
        &(r \in \pi) 
        \\
        &\leq d_G(s, t) + 2W_{\max} 
    \end{aligned}\end{equation*}

    Finally, we compute the running time. 
    At filter-edge step, we run the shortest-path-tree algorithm for each vertex in $S$, thus it is $\tilde{O}(m|S|) =\tilde{O}(m\frac{n}{\mathcal{E}})$ time in expectation. After that, $G$ has $m' \leq n\mathcal{E}$ edges left. $d$-light initialization runs in $O(m')$ time. The running time of MECSP is $\tilde{O}(gm')$ for a source vertex where $g$ is the maximum number of missing edges. Thus, adding paths from all vertices in $D_i$ to $R$ costs $\tilde{O}(|D_i|2^im') = \tilde{O}(\frac{n}{2^i}2^im') = \tilde{O}(\frac{nm'}{d})$ in expectation. 
    We have $\log(n)$ different sets $D_i$, therefore the expected running time of adding paths from all vertices of these sets to $R$ is $\tilde{O}(\frac{nm'}{d}$).
    The total running time in expectation
    is $\tilde{O}(\frac{mn}{\mathcal{E}} + \frac{nm'}{d})$. 
    Substitute $m' = n\mathcal{E}$, $d = n^{1/3}, \mathcal{E} = n^{2/3}$, it is $\tilde{O}(mn^{1/3} + n^{7/3}) = \tilde{O}(n^{7/3})$.
    \end{proof}

    \subsection{\texorpdfstring{$+(6+\epsilon)W_{\max}$ spanners of $\tilde{O}(\frac{n^{1/3}}{\epsilon})$ edges in $\tilde{O}(n^2)$ time}{\texttwoinferior}}	
    \label{sec:detail_fast1}
    
    This section includes details of Item 1 of \Cref{thm:time-6W}. 
    We apply the fast construction of Woodruff~\cite{Woo10} for $+6$ spanners, but
    replace $d$-initialization with $d$-light initialization and plug the subroutine Weak CSSSP instead of FHEASW of Woodruff~\cite{Woo10}.
    Recall that FHEASW is the subroutine to find paths with a constraint on missing edges in unweighted graphs, and can not be extended directly to weighted graphs, while CSSSP is applying Dijkstra in edge-reweighted graphs.

    \begin{algorithm}
    \caption{$+(6+\epsilon)W_{\max}$ spanners of $\tilde{O}(\frac{n^{4/3}}{\epsilon})$ edges in $\tilde{O}(n^2)$ time}\label{alg:6eWmax_Fast}
    \begin{algorithmic}[1]     
    \State Given a graph $G$ and a parameter $\varepsilon$
    \State Let $H$ be a $d$-light initialization of $G$ where $d = n^{1/3}$.
    \State Create $R$ as a $p_r$-random sample where $p_r = \frac{2\ln{n}}{d}$. 
    
	\For {$j$ from $\log(n)$ down to $0$}
    \For {$i$ from $\log(n)$ down to $0$}

    \If {$2^j > 2^id$}
        {Assign $p_{ji} = \frac{2\ln{n}}{2^j}$. }
    \Else { Assign $p_{ji} = \frac{\ln{n}}{2^id}$.}  
    \EndIf
    \State Create $D_{ji}$ by randomly sampling vertices with probability $p_{ji}$.
    \For{$v \in D_{ji}$} 
    \State Apply Weak CSSSP with source path $v$, $g = 2^{i+1}$ and $\epsilon_0 = \epsilon/2$. 
    \State Add to $H$ the shortest path returned by Weak CSSSP from $v$ to all vertices $u \in R$.    
    \EndFor
    \EndFor
    \State Remove from $G$ all edges of vertices whose degree is at least $2^{j}$.
    \EndFor
    \end{algorithmic}
    \end{algorithm}

	\begin{theorem}
        \Cref{alg:6eWmax_Fast} returns a $+(6 + \epsilon) W_{\max}$ spanner of $\tilde{O}(\frac{n^{4/3}}{\epsilon})$ edges in $\tilde{O}(n^2)$ time. 
	\end{theorem}

    \begin{proof}
    Firstly we analyze the stretch of shortest paths.
    Consider the shortest path $\pi$ of a pair $(s, t)$.
    If $\pi$ does not have a missing edge after $d$-light initialization, it is preserved in $H$.
    Otherwise,
    let $s', t'$ be endpoints of missing edges that are respectively closest to $s$ and $t$. 
    $s', t'$ are $d$-heavy vertices, thus
    by~\Cref{lmm:vertexhit}, there exists $u_1, u_2 \in R$ such that $(s', u_1), (t', u_2)$ are in $H$ and weights of these edges are at most $W_{st}$.
    Suppose that $\pi$ has $l \in [2^i,2^{i+1})$ missing edges and its highest degree of vertices is $d^* \in [2^j, 2^{j+1})$. 
    If $d^* \geq 2^j > 2^id$, then there is a $2^j$-heavy vertex $r$ in $\pi$, and it is hit by a vertex $v \in D_{ji}$ when $p_{ji} = \frac{2\ln{n}}{2^j}$ by~\Cref{lmm:vertexhit}.
    Otherwise, by~\Cref{lmm:pathhit}, the set $D_{ji}$ with probability $p_{ji} = \frac{\ln{n}}{2^id}$ has a vertex $v$ that hits the path $\pi$ at $r$. 
    Therefore, in both cases, there is an edge $(v, r)$ in $H$ where $v \in D_{ji}$ and $r\in \pi$.
    By applying Weak CSSP for every $u\in D_{ji}$ to all vertices in $R$, there are paths from $v$ to $u_1, u_2$ added to $H$.
        
    Let $\pi'$ be a path from $u_1$ to $v$ in $G$ by walking on $(u_1, s')$, $\pi(s', r)$ and $(r, v)$.
    We remove only edges of vertices whose degree
    is at least $2^{j+1}$, while the highest degree of $\pi'$ is $d^* < 2^{j+1}$, thus all edges of $\pi'$ are in $G$. Observe that $|\pi'|  \leq d_G(s', r) + W_{st} + W_{\max}$ and $\pi'$ has less than $2^{i+1}$ missing edges. 
    Therefore, setting $g = 2^{i+1}$ and $\epsilon_0 = \epsilon/2$, by \Cref{lmm:wcssp}, the subroutine returns a path $\pi^*$ going from $v$ to $u_1$ such that 
    $\pi^*$ has at most $6g/\epsilon$ missing edges and $|\pi^*| \leq d_G(s', r) + W_{st} + (1 + \frac{\epsilon}{2})W_{\max}$. 
    Similarly, the subroutine returns $\pi^{**}$ going from $v$ to $u_2$ such that its number of missing edges is at most $6g/\epsilon$, and $|\pi^{**}| \leq d_G(r, t') + W_{st} + (1+ \frac{\epsilon}{2})W_{\max}$.
    The algorithm adds $\pi^*$ and $\pi^{**}$ to $H$, thus by triangle inequality, we have:
    \begin{equation*}\begin{aligned}   
    d_H(u_1, u_2) &\leq d_H(u_1, v) + d_H(v, u_2) \\
    &\leq |\pi^*| + |\pi^{**}|\\
    &\leq d_G(s', r) + d_G(r, t') + 2W_{st} + (2+\varepsilon)W_{\max}     \end{aligned}\end{equation*}
    Since $r \in \pi(s', t')$, we obtain $d_H(u_1, u_2) \leq d_G(s', t') + 2W_{st} + (2+\epsilon)W_{\max}$.  
    By definition of $s'$ and $t'$, $d_H(s, s') = d_G(s, s')$ and $d_H(t, t') = d_G(t, t')$. Therefore: 
	\begin{equation*}\begin{aligned}
        d_H(s, t) &\leq  d_H(s, s') + w(u_1, s') + d_H(u_1, u_2) + w(u_2, t') + d_H(t', t) \\
       &\leq d_G(s, s') + W_{st} + \left(d_G(s', t') + 2W_{st} + (2+\epsilon)W_{\max}\right) +  W_{st} + d_G(t', t) \\
       &\leq d_G(s, t) + 4W_{st} + (2+\epsilon) W_{\max} \\
       &\leq  d_G(s, t) + (6+\epsilon) W_{\max} 
    \end{aligned}\end{equation*}
    
    Consider the size of $H$.
    $d$-light initialization requires $O(nd)$ edges.
    When $2^j \leq 2^{i}d$, $p_{ji}=\frac{\ln{n}}{2^{i}d}$, thus 
    $E[|D_{ji}|] = \tilde{O}(\frac{n}{2^id})$ vertices. 
    In case of $2^j > 2^{i}d$, $p_{ji}=\frac{2\ln{n}}{2^j}$, thus $E[|D_{ji}|] = \tilde{O}(\frac{n}{2^j}) \leq \tilde{O}(\frac{n}{2^id})$.
    Therefore, the algorithm adds $O\left(|R||D_{ji}|\frac{6}{\epsilon}2^{i+1}\right) = \tilde{O}(\frac{n^2}{\epsilon d^2})$ missing edges for each pair $(j, i)$.
    Since we have totally $\log^2(n))$ pairs, we add to $H$ $\tilde{O}(\frac{n^2}{\epsilon d^2})$ edges in expectation.
    The expected size of $H$ is $\tilde{O}(nd+ \frac{n^2}{\eps d^2})$. With $d = n^{1/3}$, it is $\tilde{O}(\frac{n^{4/3}}{\eps})$ edges.
	
	Now we prove the running time. $d$-light initialization runs in $O(m)$. Weak CSSSP runs in $\tilde{O}(m_j)$ where $m_j$ is the number of edges of the graph at step $j$.
	Because we apply the subroutine for all vertices in $D_{ji}$, the cost for each pair $(j, i)$ is $\tilde{O}(|D_{ji}|m_j)$.
    When $2^j\leq 2^id$, 
	the highest degree of the graph is less than $2^{j+1} \leq 2^{i+1}d$, thus $m_j < nd2^{i+1}$ and  the running time in expectation is $\tilde{O}(\frac{n}{2^id}nd2^{i+1}) = \tilde{O}(n^2)$.
	In case of $2^j > 2^id$, we have $E[|D_{ji}|] = \tilde{O}(\frac{n}{2^j})$. In addition, the highest degree of the graph is less than $2^{j+1}$, therefore $m_j < n2^{j+1}$. We obtain the running time for each pair $(j, i)$ is $\tilde{O}(\frac{n}{2^j}n2^{j+1}) = \tilde{O}(n^2)$ in expectation. With $\log^2(n)$ pairs, the expected running time is $\tilde{O}(n^2)$.   In conclusion, the expected running time of the algorithm in $\tilde{O}(n^2)$.   
    \end{proof}

    \subsection{ \texorpdfstring{$+\max\{4W, 2W_{\max}\}$ spanners in $\tilde{O}(n^{12/5})$ time}{\texttwoinferior}}	\label{sec:detail_fast4}

    In this section, we speed up the Chechik's construction of $+4$ spanners~\cite{Che13} to obtain the Item 4 of \Cref{thm:time-6W}.
    We simply use MECSP to compute $\mathcal{P}_{uv}$ and add a pre-processing step that filters edges of high-degree vertices.
    After that, we extend the algorithm to construct $+4W_{\max}$ spanners by replacing the $d$-initialization with $d$-light initialization. 
    We show the algorithm for $+4W_{\max}$ only, but one can set the weights of edges as $1$ to obtain $+4$ spanners in unweighted graphs.   

    \begin{algorithm}
    \caption{Fast construction for $+\max\{4W, 2W_{\max}\}$ spanners}\label{alg:4w_fast}
     $\mathcal{E} \gets n^{3/5}$,
        $d \gets n^{2/5}$,
        $l \gets n^{1/5}$.
    \begin{enumerate}
        \item Filtering edges:

        $S$ $\gets$ randomly sampling vertices with probability $\tilde{O}(\frac{1}{\mathcal{E}})$.
        
        Add to $H$ the shortest-path tree from each vertex of $S$ to all vertices of $V$. 
        
        Remove all edges of $\mathcal{E}$-heavy vertices out of $G$.

        \item Initialization:
        
        $H$ $\gets$ $d$-light initialization of $G$
        
        \item Randomly sampling for paths with at least $l$ missing edges:
        
        $R$ $\gets$ randomly sampling vertices with probability $\tilde{O}(\frac{1}{dl})$. 
        
        Add shortest-path tree from each vertex of $R$ to all vertices of $V$.
        
        \item Randomly sampling for paths with less than $l$ missing edges:
        
        $D$ $\gets$ randomly sampling vertices with probability $\tilde{O}(\frac{1}{d})$. 

        For each $u \in D$: Apply $\Call{MECSP}{G, u, l}$. 
        Add to $H$ paths returned by MECSP from $u$ to every $v \in D$
    \end{enumerate}        
    \end{algorithm}
    
	\begin{theorem}
	\Cref{alg:4w_fast} returns a $+\max\{4W, 2W_{\max}\}$ spanner of $\tilde{O}(n^{7/5})$ edges of $G$ in $\tilde{O}(n^{12/5})$ time.
	\end{theorem}
 
	\begin{proof}

    Consider a shortest path $\pi$ going from $s$ to $t$ in $G$. If all of its edges are added by $d$-light initialization then $d_H(s, t) = d_G(s, t)$. If existing a $\mathcal{E}$-heavy vertex $r$ in $\pi$, by the first item of~\Cref{lmm:rdm_smpl}, $N[S]\cup \{r\}\neq \emptyset$, which means there exists $(u, r)$ in $G$ such that $u \in S$. Thus, there is a path walking from $s$ to $u$ and $u$ to $t$ in $H$:  
    \begin{equation*}\begin{aligned}
        d_H(s, t) &\leq d_H(s, u) + d_H(u, t) &\textrm{(triangle inequality)} \\
        &= d_G(s, u) + d_G(u, t) &\textrm{(Add shortest-path tree from $u$ to $V$)} \\
        &\leq  d_G(s, u) + w(r, u) + d_G(r, t) + w(r, u) &\textrm{(triangle inequality)} \\
        &= d_G(s, t) + 2w(r, v) &r \in \pi(s, t) \\
        &\leq d_G(s, t) + 2W_{\max} 
    \end{aligned}\end{equation*}   
    
    If $\pi$ is not preserved by these cases, there exists a missing edge and $d$-heavy vertices. When $\pi$ has at least $l$ missing edges, we consider step 3 of the algorithm. 
    By \Cref{lmm:pathhit}, there is an edge $(r, v)$ such that $v \in R, r \in \pi$ and $w(r, v) \leq W_{st}$. Therefore:
    \begin{equation*}\begin{aligned}
        d_H(s, t) &\leq d_H(s, v) + d_H(v, t) &\textrm{(triangle inequality)} \\
        &= d_G(s, v) + d_G(v, t) &\textrm{(Add shortest-path tree from $v$ to $V$)} \\
        &\leq  d_G(s, r) + w(r, v) + d_G(r, t) + w(r, v) &\textrm{(triangle inequality)} \\
        &= d_G(s, t) + 2w(r, v) &r \in \pi(s, t) \\
        &\leq d_G(s, t) + 2W_{st} 
    \end{aligned}\end{equation*}

    Now consider the case that $\pi$ has less than $l$ missing edges and step 4. Let $s', t'$ be missing-edge endpoints that are respectively closest to $s$ and $t$. 
    It follows that $s', t'$ are $d$-heavy vertices.
    By \Cref{lmm:vertexhit}, there are $(s', u_1), (t', u_2)$ in $H$ such that $u_1, u_2 \in D$ and $w(s', u_1) \leq W_{st}, w(t', u_2) \leq W_{st}$. 
    Let $\pi'$ be the path from $u_1$ to $u_2$ by walking on $(u_1, s')$, $\pi(s', t')$ and $(t', u_2)$, i.e. $\pi' = (u_1, s') \odot \pi(s', t') \odot (t', u_2)$. Observe that $\pi'$ has less than $l$  missing edges, thus $\pi'$ is considered to find a path between $u_1$ and $u_2$. It implies $d_H(u_1, u_2) \leq |\pi'| \leq d_G(s', t') + 2W_{st}$. Then:
    \begin{equation*}\begin{aligned}
        d_H(s, t) &\leq d_H(s, s') + w(s', u_1) + d_H(u_1, u_2) + w(u_2, t') + d_H(t', t) &\textrm{(triangle inequality)} \\
        &\leq d_G(s, s') + W_{st} + (d_G(s', t') + 2W_{st}) + W_{st} + d_G(t', t) \\
        &\leq d_G(s, t) + 4W_{st} 
    \end{aligned}\end{equation*}

    To prove the size of $H$, we compute the amount of edges added step-by-step. Filter-edge step requires $O(|S|n) = \tilde{O}(\frac{n^2}{\mathcal{E}})$ edges in expectation. Step 2 and 3 add respectively $O(nd)$ and $O(|R|n)=\tilde{O}(\frac{n^2}{dl})$ edges. The last step adds $O(|D|^2l) = \tilde{O}(\frac{n^2}{d^2}l)$. With $\mathcal{E} = n^{3/5}, d = n^{2/5}, l = n^{1/5}$, we obtain the expected size of $H$ is $\tilde{O}(n^{7/5})$.  
    
    Finally, we compute the running time. 
    In the first step, we run the shortest-path-tree algorithm for each vertex in $S$, thus it is $\tilde{O}(m|S|) =\tilde{O}(\frac{mn}{\mathcal{E}})$.
    After filtering, $G$ has $m' \leq n\mathcal{E}$ edges left. 
    Thus the running time of step 2 and step 3  are respectively $O(m')$ and $\tilde{O}(m'|R|)=\tilde{O}(m'\frac{n}{dl})$.
    The final step runs MECSP for each vertex in $D$, therefore it costs $\tilde{O}(m'l|D|)=\tilde{O}(m'\frac{nl}{d})$.
    Since $m' \leq n\mathcal{E}$, the expected running time is $\tilde{O}(m + \frac{mn}{\mathcal{E}} + \frac{n^2\mathcal{E}l}{d}) = \tilde{O}(\frac{mn}{\mathcal{E}} + \frac{n^2\mathcal{E}l}{d})$, which is $\tilde{O}(mn^{2/5} + n^{12/5}) = \tilde{O}(n^{12/5})$.
    \end{proof}

    \appendix
    \section{Appendix}
    \label{sec:appendix}

    \randomsample*
    
    \begin{proof}
    With the first item, given a $d$-heavy vertex $u$, we have:
    \begin{equation*}
        \begin{aligned}
            P[N[u]\cap S = \emptyset] = \left(1 - \frac{2\ln{n}}{d}\right)^{|N[u]|} \leq \left(1 - \frac{2\ln{n}}{d}\right)^d = \frac{1}{n^2} 
        \end{aligned}
    \end{equation*}
    By union bound, this probability for every vertex in $G$ is at most $\frac{1}{n}$.

    Similarly, we prove the second item:
    \begin{equation*}
        \begin{aligned}
            P[N[\pi]\cap S = \emptyset] = \left(1 - \frac{\ln{n}}{k}\right)^{|N[\pi]|} \leq \left(1 - \frac{\ln{n}}{k}\right)^k = \frac{1}{n} 
        \end{aligned}
    \end{equation*}
    \end{proof}

 \vertexhit*
    
    \begin{proof}
    Create $D$ by randomly selecting each vertex with probability $ \frac{2\ln{n}}{d}$.
    Given a $d$-heavy vertex $u$ in $\pi$, let $S(u) = \{v \in N(u) \mid (u, v) \textrm{ is in $d$-lightest edges of } u\}$.  
    Because of $d$-light initialization, edges $(u, v)$ for $v \in S(u)$ are added to $H$.
    Observe that their weights are at most the weight of a missing edge of $u$ in $\pi$.
    It implies that these weights are at most $W_{st}$.
    We have:
    \begin{equation*}
        P[S(u) \cap D = \emptyset] = (1 - p_D)^{|S(u)|} 
        = \left(1 - \dfrac{2\ln{n}}{d}\right)^{d} = \dfrac{1}{n^2}        
    \end{equation*}
    By union bound, this probability for every vertex $u \in \pi$ is at most $\dfrac{1}{n}$.         
    \end{proof}

 \paragraph{Acknowledgement.~} Hung Le was supported by the NSF CAREER Award No.\ CCF-2237288 and an NSF Grant No.\ CCF-2121952. An La was supported by the NSF CAREER Award No.\ CCF-2237288, an NSF Grant No.\ CCF-2121952, and NSF TRIPODS  CISE-1934846.
\bibliography{ref}

\end{document}